# AC Loss in Striped (Filamentary) YBCO Coated Conductors Leading to Designs for High Frequencies and Field-Sweep Amplitudes

M.D. Sumption[1], E.W. Collings[1], and P.N. Barnes[2]

[1] Laboratories for Applied Superconductivity and Magnetism, Materials Science and Engineering Dept., The Ohio State University, OH 43210, USA

[2] Propulsion Directorate, Air Force Research Laboratory, Wright-Patterson AFB, OH 45433, USA

*Abstract*

AC losses of YBCO coated conductors are investigated by calculation and experiment for the higher frequency regime. Previous research using YBCO film deposited onto single-crystal substrates demonstrated the effectiveness "striping" or filamentary subdivision as a technique for AC loss reduction. As a result of these studies the idea of subdividing YBCO "coated conductors" (both YBCO, overlayer, and even underlayer) into such stripes suggested itself. The suggestion was implemented by burning grooves into samples of coated conductor using laser micromachining. Various machining parameters were investigated, and the striping and slicing characteristics are presented. Loss measurements were performed on unstriped as well as striped samples by the pick-up coil technique at frequencies of from 50-200 Hz at field sweep amplitudes of up to 150 mT. The effect of soft ferromagnetic Fe shielding was also investigated. The results of the experiments form a starting point for a more general study of reduced-loss coated conductor design (including hysteretic-, coupling-, normal-eddy-current-, and transport losses) projected into higher ranges of frequency and field-sweep amplitude with transformer and all-cryogenic-motor/generator applications in mind.

**Key Words: Conductor design, AC loss, hysteretic loss, coupling loss, eddy current loss, coated conductor, striping, magnetic shielding, YBCO, IBAD process, RABiTS process**

# 1. INTRODUCTION

## 1.1 Background

Advances in the processing of multifilamentary (MF) superconducting strand particularly during the late 1960s and early 1970s [1] were accompanied by a burgeoning interest in large-scale applications of superconductivity [2]. Both DC- (high field magnets for particle acceleration, magneto-hydrodynamic-based power, fusion, etc) and AC applications were being engineered. Among the AC applications being considered worldwide were: generators, power transmission lines, transformers, and motors. Accompanying the development of AC applications the study of AC loss in superconductors was being actively pursued, particularly by a group at the then Westinghouse Research and Development Center. With MF composite strand as a model system in some cases, the Westinghouse group analyzed loss due to: (i) an external AC magnetic field, (ii) the self field of an AC transport current, (iii) AC field and DC current, (iv) AC field and AC current. Although based on low temperature superconductors (LTS) at the time the theoretical findings of that group (as detailed by Carr in the first edition of his book [3] and elsewhere) are closely adaptable to the AC loss problems of high temperature superconductors (HTS) as explained in the second edition of Carr's book [4].

Now, just as before, the emerging availability of high performance strand is again stimulating the design and eventual construction of superconducting machinery but this time with HTS superconductors. Reaching maturity as HTS conductors are the "first generation" Ag-clad powder-in-tube composites based on Bi(Pb)-Sr-Ca-Cu-O referred to as Bi:2212/Ag and Bi2223/Ag. Under development and also being considered for energy and power applications are the "second generation" HTS coated conductors based on epitaxial $YBa_2Cu_3O_x$ (Y:123, or YBCO). The particular conductor design goal being approached here is based on a US Air Force (USAF) desire for a high frequency ($f$), high field amplitude ($H_m$), generator with HTS field- and stator windings. The need for hundreds of Hz (up to 500 Hz), a $\mu_0H_m = 1$ T, and acceptable AC loss/m of conductor pushes the properties of striped coated conductor to the very limit of attainability.

## 1.2 The Coated Conductor

The coated conductor is a multilayer composite tape in which textured YBCO is supported on a strong metallic underlayer and protected in most cases by a Ag overlayer. Buffer layers, typically oxides, are sandwiched between the YBCO and substrate. Three major processing methods are in use, their designations referring to the techniques used to texture, or apply texture to, the underlayer creating an epitaxial template for the subsequent growth of biaxially textured YBCO layer. They are known as are the rolling-assisted-biaxially-textured-substrate (RABiTS) method, the ion-beam-assisted-deposition (IBAD) method, and the inclined-substrate-deposition (ISD) method [5]. In the RABiTS approach the biaxially textured template is induced in a metallic foil, typically a Ni-alloy, by a series of deformations followed by a heat treatment. With IBAD and ISD the epitaxial template is created in an initial oxide buffer layer by controlled depositions using either ions (IBAD) or angular deposition (ISD) to induce the texture. Numerous oxides are available for buffering including $CeO_2$, YSZ, MgO, GdZO, $SrRuO_3$, $LaNiO_3$, and $(La,Sr)MnO_3$. The final tape is generally coated with a protective layer of Ag, and a Cu layer may be added for stabilization.

### 1.3 Conductors for AC Applications

There is a need for AC-tolerant YBCO coated conductor for 60 to 500 Hz applications. Under such conditions, especially the higher frequency end of the spectrum, it will be essential to determine and minimize AC losses. AC magnetic field losses are greatest when the field is normal to the wide face of the tape (face-on, FO) and least in when it is parallel to that face (edge-on, EO). In practice it will generally not be possible to engineer the pure EO condition. Some FO component of the applied field (the so-called "stray field") will always be present and will call for special conductor designs to minimize that loss. Stray field loss components are expected for transformers and inductors, as well as all-superconducting generators and motors, which represent extreme examples.

To combat FO loss in general the conductor is subdivided into $N_f$ parallel filaments running down the length of the tape. To prevent magnetic re-coupling of these filaments it is necessary to limit the effective length over which these filaments are exposed to a field of a given direction. With round multifilamentary (MF) strands, twisting is routinely applied. But for coated conductors, which are high aspect ratio tapes, the feasibility of twisting is questionable. For tape there are physical limits to twist pitch reduction. But even if twisting is achievable the tape would assume a helix-like profile, occupying a large volume and severely reducing the engineering critical current density. Even so it is important to recognize that some kind of twisting or transposition will be needed in a conductor of finite length with its required low resistance current contacts at each end. In practice the twisting, if not performed on the conductor itself as part of the processing sequence (as in round LTS MF strands) may occur naturally during the winding of the superconducting machine. To control loss both filamentary subdivision and twisting are needed. In this study, however, we focus on the filamentary subdivision of coated conductor and its effectiveness in loss reduction.

### 1.4 Organization of the Paper

We begin the paper by referring to some of our previous studies of striped YBCO films in which the effects of filamentary subdivision on hysteretic loss were studied using vibrating sample magnetometry (VSM) at 4.2 K in slow ramp-rate fields of amplitudes as high as 9 T, and in which insights were developed concerning the loss-controlling properties of *stripe width* and *film thickness*. We then move on to describe the striping of some YBCO coated conductor samples and a series of measurements performed inductively on them using pick-up-coil magnetometry at 4.2 K in AC magnetic fields of frequencies ranging from 50 to 200 Hz and amplitudes of up to 150 mT. The effects of ferromagnetic partial shielding are also described. Then recognizing that these measurements: (1) are just part of a planned more extensive research program involving much higher frequencies and fields (2) point the way towards the design of low loss conductors for machine applications, we use the results as a starting point for a design study of low loss coated conductors for use over a range of frequencies of up to 500 Hz and field amplitudes of up to 1 T. The collected results of the design study are tabulated and discussed, and conclusions are offered.

## 2. EXPERIMENTAL

### 2.1 Patterning (Striping) of YBCO Films

As mentioned above filamentary subdivision is routinely applied to LTS round strands in order to suppress flux jumping (under all conditions) and hysteretic loss (under AC field conditions). Filamentary subdivision of first generation HTS strands (which become fully bridged after reaction heat treatment) is applied for entirely different reasons – crack suppression and increase of the Ag/HTS interface surface area. With second generation HTS (coated) conductors we return to one of the original motivations – if not flux jump suppression, then certainly AC loss reduction. Carr has been a long time contributor to the theory of loss reduction by filamentary subdivision, first in the context of LTS conductors [3] and later with regard to second generation HTS [4,12].

Numerous techniques have been used to pattern YBCO, including wet chemical etching, ion beam etching, and laser ablation. In some cases the aim is to completely remove material, in others to merely degrade the superconductivity of the stripe. Laser ablation has been used to form stripes in some cases by totally removing material, in other cases to cause it to react in an oxygen-rich environment to degrade the SC properties. Other methods used have been: (i) ion implantation [6] or (ii) inhibition- or reactive-patterning [7] which locally degrades the properties of the YBCO film by destroying its crystalline structure or, while retaining the structure, lowers its electrical conductivity [6,7]. In this latter case stripe patterns made were typically 10 μm wide, with interface transitions of about 0.1 μm. Another striping technique uses masking and Ar-ion-beam etching [8]. These earlier studies of YBCO patterning have been made in the pursuit of electronic applications. But with regard to "large-scale applications" some recent studies have focused on the use of the striping of conductors as a way of reducing their AC losses.

We have demonstrated the utility of patterning or striping the conductor (at least from a short-sample standpoint) using YBCO films deposited on single crystal substrates [9, 10]. The striping was performed in one case by laser ablation, where the already deposited YBCO was fully removed (ablated) from the target interfilamentary zones. In this work the results of $T_c$ and DC susceptibility measurements performed on the striped YBCO films revealed almost no degradation in the filamentary areas remaining. Magnetization (M-H) loops were measured and their areas shown to follow the expected dependence on filamentary width. Furthermore, detailed analysis of the results indicated that the proportionality of loss to stripe width was superimposed on a general reduction of loss in response to decrease in stripe *thickness* under appropriate conditions. Filaments have also been formed mechanically [11] and the expected loss reductions seen.

Based in large part on the insights developed in studying striped YBCO/LaAlO coupons we went on to pattern several samples of YBCO coated conductor and study their AC loss properties.

## 2.2 Sample Preparation

### 2.2.1 Conductor and Patterning

The coated conductor specimens were fabricated externally. The YBCO layer was an IBAD-deposited buffer layer on a Hastelloy-type ribbon (the substrate, or underlayer). Sputtering was used to deposit an overlayer of Ag to a thickness of 3-4 μm. The finished tapes, 0.125 mm thick and 1 cm wide were then sent to Mound Laser Photonics Inc for patterning (or striping) by laser micromachining (following the work of the Air Force Research Laboratory (AFRL) on the patterning of YBCO/LaAlO coupons [9]). The striping was done using the third harmonic of a $Nd:YVO_4$ diode-pumped solid-state laser at 355 nm whose beam was focused using a simple lens. The coated conductor, mounted on a vacuum stage attached to a translation stage, was positioned at the focal plane of the lens. The stripe was machined by translating the conductor through the focused beam. Typically, the incident average power was less than 1 W and the translation (scan) rate was 10 mm/sec. Currently, the attainable stripe and slice widths are approximately 25-30 μm.

The desired structure is shown schematically in Figure 1, where the patterning is seen to separate not only the YBCO layer but also Ag overlayer into stripes. Figure 2 shows the actual results of the laser patterning on a sample of coated conductor. The top micrograph shows that in this particular case the tape was patterned with several groove width and depth values, achieved by proper choice of laser power and scan speed. The lower figure shows a detail in the vicinity of a stripe. Figure 3 is taken looking down on the same set of experimental stripes (intentionally made with varying widths), it depicts one of the two "test" samples which were made to determine the proper laser striping and slicing parameters (here we use stripe to denote cutting into through the filaments, and slicing to denote cutting very deeply into, and in many cases all the way through, the sample). Table 1 describes the (more complete) second set of striping tests which were performed. Different power levels and scan speeds were used, resulting in various stripe depths. Additionally, either Ar or air was passed over the sample during laser processing to clean away debris. We also investigated slicing all of the way through the samples, these experiments are described in Table 2. Figure 4 (from Table 1) displays groove widths and depths as functions of scan rate and beam power. Here we see a very sensible trend for the depth of the stripe from the 0.5 W beam in air, plotted as a function of scan speed (i.e, depth decreasing with increasing scan speed). This trend is still present, although not as smooth, for the 0.75 W beam, and the 0.5 W beam in Ar. It should be noted at this point that the depths recorded in Figure 4 are the final stripe depth, and not the depth of the initial melt. Inspection of the SEM micrographs suggest that much of the time for the 0.75 W beam there is a difference in these two depths, this may also induce scatter in the data. Such effects are probably also present to a lesser degree in the 0.5 W sample. This is probably also responsible for the differences seen at 0.5 W for Ar vs air environments for the samples. Figure 5 (from Table 2) shows the slice depth and width for various numbers of passes. After the initial series of experiments, a nominal stripe depth of 25 μm was chosen to guarantee filament separation and all samples measured for AC loss had stripes of that depth.

### 2.2.2 Ferromagnetic Foils

In preparation for "shielding effect" measurements Fe strips were incorporated into some of the YBCO coated conductor sample packs, Table 3. The Fe strips were 99.9975% pure, stock number 10866 from Alfa Aesar, and had dimensions 0.025 x 10.0 x 50.0 mm.

## 2.3 Measurements

AC loss was measured using a pick-up coil magnetometer immersed in boiling liquid nitrogen. The field was applied by a solenoidal primary with a clear bore of 76 mm and a length of 186 mm. It was wound with 1510 turns of 1 mm OD Cu wire. The applied frequency ranged from 50 to 200 Hz with a maximum amplitude of 150 mT. The pick-up coil consisted of two single-layer racetrack-like windings (6 cm x 1 cm) each with 238 turns. Wire diameter was 100 $\mu$m and the inductance of the coil was 0.98 mH. The compensating coil was nominally matched to the pick-up coil, the two being placed at positions symmetrically displaced from the center of the primary coil (along the axial direction of the solenoid). The pickup coils were designed for 5-6 cm long samples. The system is configured to take the data as $M$-$H$ loops which were then integrated to find the total loss. Calibration was performed in terms of the magnetization of high purity Fe as determined by measurement on a sample of the same material in a calibrated VSM. The VSM was calibrated at higher fields using Ni spheres and cylinders. Sample packs assembled for measurement are listed in Table 3.

## 3. Results

### 3.1 Unmodified Coated Conductor

Figure 6 depicts a typical set of $M$-$H$ loops, in this case for Sample Pack A (Table 3) which consists of 6 unstriped YBCO tapes, each 5 cm long, stacked on top of one another. The measurements were performed at 77 K and, in this case, under 50 Hz applied fields of up to 150 mT in amplitude oriented perpendicular to the tape faces (the FO orientation). The magnetic moment, $M$, expressed in SI units per m leads directly to statements of the AC loss in the useful units J/m per cycle and hence W/m or subsequently W/kAm if desired. The $M$-$H$ loop is featureless with no evidence of flux jumping.

### 3.2 Fe/YBCO Multilayer Results

Magnetometer measurements under the above conditions were made on samples surrounded by Fe strips, Figure 7. Four strips of Fe each having about the same surface dimensions as the YBCO were placed two above and two below the central YBCO pair to form Sample Pack B (Table 3). This is probably not an optimal configuration for shielding, but does in fact show some suppression of loss for lower field amplitudes. The Fe response tends to rotate and change the character of the loops – the former due to the simple addition of Fe magnetization, the latter to influences of the Fe shielding on field distributions and thus YBCO magnetization.

### 3.3 Comparisons of Striped and Non-Striped Conductors, with and without Fe Multilayers

Magnetometer measurements under the above conditions, and with and without the included Fe layers, were made on coated conductor tapes that had been striped by laser into ten filaments. These were Sample Packs C and D (Table 3). The results, expressed in the format mJ/m per cycle, are displayed in Figures 8 and 9, along with data from unstriped samples (Figures 6 and 7, also integrated to obtain the total per-cycle loss/m).

### 3.3.1 Striping without Fe

Referring to Figure 8, we notice that the striping into 10 filaments reduces the loss by a factor of about 10, as expected from the standard expression for per-cycle hysteretic loss – see Eqn.(1) below (which of course is valid only for $H_m$s well above the penetration field, $H_p$). We notice also that striping is uniformly effective at all field-sweep amplitudes, $H_m$. This agrees with the principle that loss reduction scales with filament-width reduction at *all* $H_m$ -- even as $H_m$ approaches $H_p$, according to arguments associated with Eqns. (2) and (3) below. Figure 9 displays the per-cycle loss (at fixed $H_m = 150$ mT) as function of frequency. The per-cycle hysteretic loss of the YBCO is of course independent of frequency – Eqn.(1) below. The very slight frequency dependency observed in the striped tape is presumably coming from eddy currents in the Hastelloy underlayer with perhaps a small contribution from the Ag overlayer (depending on the strength of its electrical connectivity to the YBCO layer

### 3.3.2 Striping with Fe

We notice in Figure 8 that at 50 Hz, the presence of the Fe reduces the loss of the unstriped sample, but increases it for the striped sample. When Fe strips are incorporated into the conductor pack three mechanisms operate to influence the total loss: (1) eddy currents ($f^2$-dependent); (2) hysteretic loss ($f$-dependent); (3) partial shielding of the YBCO (ineffective at saturation field strengths). Differences in their internal layouts (Table 3) complicate a direct comparison between the Fe effects in the striped and un-striped sample packs. Nevertheless some useful observations can be made: With regard to striped-conductor loss, Figures 8 and 9 suggest that any beneficial magnetic influence of the Fe (shielding or field tailoring) is masked by loss due to its eddy currents and the fact that the striped YBCO loss is relatively small to begin with. On the other hand, when the YBCO loss is large, as in the unstriped conductor, the presence of Fe can reduce it slightly (see in particular Figure 8). This observed reduction is evidence that in this case loss reduction by shielding or field tailoring is greater than loss augmentation by hysteresis and eddy currents.

### 3.4 Comparison with Theory

How does the measured total loss of our coated conductor sample (dominated by the hysteretic loss of the YBCO layer) compare with standard theory? Four-terminal measurements of the 1 cm wide conductor's self-field critical current gives a $J_c$ of 2.5 x 10$^9$

A/m$^2$ (based on an $I_c$ of 50 A and a layer thickness of 2 μm). Figure 8 suggests then that $J_c$ averaged over the whole 150 mT cycle is just under ½ of this. Standard theory -- Eqn.(1) below – then predicts an energy loss per 150 mT cycle of about 38 mJ/m. As for the experiment, Figure 9 indicates an *f*-independent loss per cycle at $H_m$ = 150 mT of 25 mJ/m. The effect of striping into $N_f$, "filaments" (with negligibly thin grooves) is to divide this loss by about $N_f$ as shown in Figure 9.

### 3.5 Implications for Striped Conductor Design

Experiments performed initially on YBCO/LaAlO films set the stage for a much more detailed study of coated conductor patterning techniques. Close control of the focused laser beam enable grooves to be cut not only through the overlayer and YBCO layer but also into the underlayer (Table 1) and through it (Table 2). Measurements, first by VSM, and later using pick-up-coil magnetometery at 50-200 Hz demonstrate the effectiveness of striping in the reduction of hysteretic loss. Furthermore, since eddy current loss goes as sample width squared, the possibility of cutting the overlayer along with the YBCO, and also of cutting through the underlayer as well offers the possibility of drastically reducing the normal metal eddy current losses. With these observations, among others, as a starting point, we go on to analyze the loss components that would be operative in a composite coated conductor. In the following design study we estimate these losses first for 500 Hz/1 T (by way of example) and go on to extend the range of these estimates to other frequencies and field amplitudes, Table 4.

## 4. CONDUCTOR DESIGN FOR AC APPLICATIONS

### 4.1 Outline

After considering the relative magnitudes of applied field- and self-field losses experienced by coated conductors the present paper focusses on the dominant source of AC loss (external field) and its reduction by striping, the control of the resulting coupling component of loss, and the overall design of striped conductors. After demonstrating that for a simple coated conductor in the presence of AC transport current, the hysteretic component is the dominant source of loss, we go on to consider the reduction of that loss by striping, the resulting coupling loss mediated by resistive paths between the stripes and end contacts, and eddy current loss in the normal metal layers. Also explored is the effect of ferromagnetic shielding on the applied-field loss. In most cases we consider the losses generated by fields applied normal to the tape surface (face-on, FO, orientation) and parallel to it (edge-on, EO, orientation),

We arbitrarily set the operating temperature at 77 K and tape specifications (arbitrarily but realistically) at: critical current density, $J_c$, $10^{10}$ A/m$^2$; total width, *w*, 1 cm; stripe width, *d*, various; thickness, *t*, 2 μm. We express the 500 Hz AC losses of the tape's various components in the units W/m; although not taken here, a small additional step would enable the loss to be expressed in terms of another useful unit, W/(kA.m).

### 4.2 Flux Jump Stability

In designing a superconducting wire the first consideration has always been its flux-jump stability. A sample of superconductor of unspecified size in the virgin state will acquire sufficient energy to initiate magnetothermal instability when an applied field exceeds a critical value $H_{fj}$. However, the available energy can be metered by reducing the width of the sample "seen" by the field; thus a superconductor of finite width will never acquire enough energy to flux-jump provided the applied field reaches penetration (at $H_p$) before the estimated $H_{fj}$. Initially intended to eliminate flux-jump instability, and subsequently to control hysteretic loss, the idea of flux metering by width reduction has guided the design of MF strand since the 1960s [13]. Returning to HTS superconductors, adiabatic flux jumping in bulk YBCO has been described by several authors [14,15]. The simple criteria they developed indicate flux jump immunity for the tapes of the kind under consideration especially in the light of the thickness-suppressed $H_p$ generally exhibited by tapes [4, p.182], see below. Indeed flux jumping is completely absent from the hysteresis loops depicted in the following figures.

Of course a topic of more relevance to the design of coated conductors exposed to AC fields is the "dynamic stability of films", a subject recently treated theoretically by Mints and Brandt [16] and awaiting adaptation to the present conductors and conditions.

### 4.3 Hysteresis Loss for Tapes

### 4.3.1 The FO Orientation

The per cycle hysteretic loss per unit volume of a superconducting slab of width $w$ perpendicular to an applied sinusoidal AC field of amplitude $H_m$ is given in SI units by [4, p.182]:

$$Q_h \approx \mu_0 w J_c H_m \qquad (1)$$

where $\mu_0$ is the permeability of free space, and $J_c$ is the critical current density. This expression is applicable as long as $H_m \gg H_p$ which in the case of a slab of thickness $t$ is given by:

$$H_p \approx J_c \left( \frac{t}{\pi} \right) \left( \ln \frac{w}{t} + 1 \right) \qquad (2a)$$

or

$$H_p \approx \left( \frac{5}{2\pi} \right) H_d \left( \ln \frac{w}{t} + 1 \right) \qquad (2b)$$

in which $H_d$, the "thickness penetration field", is defined as $H_d = 0.4 J_c t$. For our model coated conductor ($J_c = 10^{10}$ A/m$^2$, $w = 10^{-2}$ m, $t = 2$ μm) $\mu_0 H_p$ is about 76 mT and $\mu_0 H_d$ is 10 mT. We conclude that in an FO field of amplitude 1 T applied perpendicular to $w$ Eqn. (1) holds and that our model conductor would experience a hysteretic power loss ($P = $ $Q_{hf}$) at 500 Hz of $5 \times 10^4$ MW/m$^3$ -- equivalent to 1.0 kW/m using the conversion factor $2 \times 10^{-8}$ m$^3$/m. The assumption $H_m \gg H_p$ will be valid for high field applications but not for small $H_m$ applications, e.g. transformers. In case $H_m < H_p$, the hysteretic loss, $Q_h$, is modified by a prefactor $N < 1$ -- a rapidly decreasing function of $H_m/H_p$, as detailed in [10], see also below. The parameter $N$ may be expressed in the form [17,18]

$$N = 2\left(\frac{H_d}{H_m}\right) Ln\left(\cosh\frac{H_m}{H_d}\right) - \tanh\left(\frac{H_m}{H_d}\right) \approx \left(1 - \frac{1.4}{(H_m/H_d)}\right) \qquad (3)$$

The behavior of $N$ in the high-$H_m$ and low-$H_m$ limits is conveniently discussed in terms of the ratio ($H_m/H_d$): Referring the exact expression for $N$, in the high-$H_m$ limit (e.g. $H_m \rightarrow 1$ T) $N \rightarrow 1$ leading to the recovery of Eqn. (1). But in the low-$H_m$ limit, $N \rightarrow 0$ as $H_m^3$, leading to $Q_h \propto H_m^4$ (as compared to the $H_m^3$ of the usual unpenetrated critical state situation).

The beneficial impact of reduced $N$ on hysteretic loss is implicit in the results of recent investigations [10][19-21]. Clearly hysteretic loss can be lowered by reducing the prefactor $N(H_m/H_d)$ to well below unity. For a given $H_m$ this can be accomplished by bringing $H_d$ (= $0.4J_ct$) closer to $H_m$ by increasing the YBCO layer thickness. With a $\mu_0H_d$ in the present case of 10 mT there is scant possibility of accomplishing this for generator stator windings with their large values of $H_m$. On the other hand, for rotor windings and transformers with FO stray fields of order 20 mT, meaningful $H_m/H_d$ reduction can be accomplished by increasing the YBCO coating thickness, $t$, beyond the 2 μm considered here. This approach is based on the assumption that the effect described remains valid when the conductor is wound into a device, in reality interactions among neighboring strands needs to be considered.

### 4.3.2 The EO Loss and Striping

In the EO orientation Eqn. (1) holds but with $w$ replaced by the strip thickness, $t$. In a strictly EO field the power dissipation would be only 200 mW/m. Of course it would be impossible to build a machine that experienced *only* EO loss. At best, the broad face of the conductor will be exposed to "stray field" from various sources including EO-field misalignment. For example a field misalignment of only 1.15° would provide an FO component equal to 2% of the "EO" main component. Such a relative stray field due to this or other sources could be expected in an air core transformer designed for mostly EO field at the windings. In terms of loss, the effect of the stray field is magnified by the aspect ratio ($w/t$) of the YBCO – 5,000 in the present example of an unstriped conductor. Thus a "2% stray field" could generate loss 100 times as great as the pure EO loss.

Clearly it is essential to reduce the conductor's FO loss, not only in response to fields deliberately directed normal to the conductor's broad face, but stemming from stray fields. The way to do this is to reduce conductor width to the micrometer level by striping. A multiplicity of such stripes, now of width $d$, is needed to maintain current carrying capacity following the fine-MF concept of hysteretic loss reduction that guided the design of composite NbTi/CuNi AC strands during the 1970s. This will be the most practical route for AC loss reduction for strand intended for EO field orientations – and it will be absolutely essential for those intended for FO orientations.

Provided the filaments are "decoupled" (see below) the hysteretic power loss per unit volume of the MF assembly is given by Eqn (1) with the symbol $w$ replaced by $d$, the width of an individual filament. Picture for example our 1 cm wide superconducting tape, 2 μm thick, subdivided into 100 stripes 50 μm wide and 50 μm apart. At 500 Hz, 1 T, this would dissipate 2.5 W/m – 400 times less than the unstriped value (not forgetting of course that the current carrying capacity, $I_c$, of the conductor has been reduced by the striping from 200 A to 100 A.). Such loss reduction is probably not yet sufficient for generator stator applications [22], but may be satisfactory for transformers. For generators, even smaller filaments are needed to control the hysteretic *loss, but beyond this other losses must be considered.*

### 4.4 Loss in the Presence of Transport Current

### 4.4.1 AC Transport Current in Zero Applied Field

In the self field of an AC transport current of amplitude $I_m$ a strip experiences a power **loss per unit length** given, according to Norris [23] (see also [4, p.192]), by

$$\left[P_{sf}\right]_L = \frac{f\mu_0 J_c (wt)^2}{\pi}\left[(1-i)\ln(1-i)+(1+i)\ln(1+i)-i^2\right]$$

$$\left[P_{sf}\right]_L = \frac{f\mu_0 J_c (wt)^2}{\pi}\left[(1-i)\ln(1-i)+(1+i)\ln(1+i)-i^2\right] \tag{4}$$

Here we leave $P$ normalized to unit conductor length, $i = I_m/I_c$ (in which the critical current , $I_c$ is 200 A for our model strip), and there is no applied field. If we pick $i = I_m/I_c$ at 0.9, and $f$ = 500 Hz, then $P$ = 1.44 W/m. Dropping the relative transport current density to 0.5 reduces this loss by a factor of 16 to 88 mW/m. Equation (4) indicates that $[P_{sf}]_L$ decreases with reduction of strip width. In fact if a tape of width $w$ is subdivided by negligibly narrow grooves into many stripes each of width $d$, the self-field loss is reduced by the factor $w/d$. The subdivision is effective only when the stripes are decoupled – which is accomplished in the presence of an applied field.

### 4.4.2 AC Transport Current in AC Applied Field

The losses are modified significantly when the tape is exposed to an AC field while carrying AC transport current. This condition has been considered by Carr [24][4 p.98], for the special cases of an AC current in-phase with a transverse field, applied to a round wire and a tape in the EO orientation. The analysis does not yet deal with an FO applied field which is the condition of most practical interest as explained above in connection with stray field effects. Nevertheless the round wire result would be closely applicable to the individual stripes of a field-uncoupled MF tape. At relatively high AC field amplitudes, $H_m \gg H_p$, the EO-tape- and round-wire results coalesce and allow us to compute the transport current contribution to the total loss using $(1/3)i^2 Q_h$ where $Q_h$ is now the FO loss for an uncoupled tape.

### 4.4.3 Combined Loss and Filamentarization

As just pointed out, for $H_m \gg H_p$, the relative combined loss, $Q_{hi}/Q_h$, for the round wire coincides with that for the EO tape. But when $H_m < H_p$, as might be the case for the ripple field experience by a rotor, Carr's analysis shows, for a given $i$, that $[Q_{hi}/Q_h]_{(wire)}$ is very much less than $[Q_{hi}/Q_h]_{(tape)}$. In this regard filamentarization has a double benefit: it not only reduces the FO $Q_h$ but allows the round-wire approximation to be applied to the AC-field/AC-current condition. However, for the present purposes, we can take the high $H_m$ limit, where the combined hysteretic and transport current loss is given by $Q_{hi} = Q_h(1+i^2/3) \equiv Q_h + Q_i$. In this case the "additive factor" due to transport current loss (when the applied field is large enough to decouple the filaments) is between 8 and 33% of the hysteretic loss, as $i$ ranges from 0.5 to 1. Thus striping, in this case, reduces both hysteretic and transport current loss.

In large $H_m$ applications the dominant loss experienced by an AC tape is the FO hysteretic component. It is essential to suppress $Q_h$ – filamentarization is the standard remedy, in this case subdivision of the superconductor into strips. But filamentarization ("striping") embodies serious problems of its own – coupling loss due to induced currents that flow along the filaments and across the normal paths between them either the intervening matrix or, in the case of fully insulated filaments, their common end current-contacts.

### 4.5 Coupling Loss

Power loss per unit volume due to coupling currents in a conductor with filamentary subdivision is given by

$$P_{coup} = \frac{1}{n\rho_{eff}} \left[ fL(\mu_0 H) \right]^2 \tag{5}$$

where $n = 2$ for a round MF strand [4, p.127] and $n = 4$ [4, p.188] for a striated flat tape. In both cases coupling loss saturates to the hysteretic loss of the monolithic conductor. In Eqn. (5) $f$, and $H_m$ have their usual meanings, and $\rho_{eff}$ is the "coupling resistivity". The coupling length, $L$, is equal to the twist pitch if the conductor is twisted relative to external fields or twice the length of an untwisted conductor. Usually coupling loss is reduced by decreasing $L$ (by twisting) and increasing $\rho_{eff}$. As suggested it can also be controlled by decreasing the "active length" of a tape segment (i.e. $L/2$) that is exposed to field of a given orientation.

### 4.6 Twisting and Related Issues

### 4.6.1 Interfilamentary Resistance

According to Eqn.(5) if sufficient $L$-reduction cannot be attained, we must look to $\rho_{eff}$ as a means of reducing coupling loss [25]. How much coupling loss is acceptable, and what is the corresponding $\rho_{eff}$? A useful estimate of it can be made by requiring that the coupling loss be equal to the hysteretic loss, which for a *twisted* tape is $(2/\pi)$ times the untwisted value, Eqn. (1) (where $w \to d$) [4, p.189]. Accordingly the "reference" $\rho_{eff}$ is given by

$$\rho_{eff} = \left(\frac{\pi}{8}\right) \frac{f\mu_0 H_m}{J_c} \left(\frac{L^2}{d}\right) \tag{6}$$

Note that this simple relationship differs from Carr's expression for $\rho_{eff}$ [4, p.191] by a factor 2.4, close enough given that the criterion has been arbitrarily chosen. With regard to limitations on $L$, present estimates are that YBCO coated conductors can support a 0.4 % tensile strain without degradation. For a tape conductor this leads to an $L \sim 30w$. If $L$ is to be 10 cm, as it might be for a Gramme ring, then $w$ must be about 1/3 cm, less than that of our reference 1 cm wide tape, but one that we will adopt for the moment. Taking as before $\mu_0 H_m = 1$ T, $f = 500$ Hz, $J_c = 10^{10}$ A/m$^2$, assuming $L = 10$ cm and a stripe width, $d$, of 50 μm, we find that $\rho_{eff} = 4 \times 10^{-6}$ Ωm (400 μΩcm). *The most important thing to note here is that this value shows that the notion of unstriped metallic layers connecting striped filaments is untenable at 500 Hz and 1 T – the loss penalty would be prohibitive.* Thus, from this point forward we will assume that the overlayer is striped along with the filaments.

It is difficult to envision a practical method of applying twist to a coated tape. As pointed out by Carr [4, p.183] Oberly's suggestion of a spiral-patterned coating on a cylindrical tube would result in a relatively large diameter conductor. Carr himself has suggested applying the twist directly to a tape bearing a striped coating [4, p.183]. Again the twisted tape will occupy much more space than the untwisted one. But spatial issues aside, the above analysis indicates that the needed $\rho_{eff}$ would be out of the range of metallic conductivity. Taken at face value Eqn.(5) indicates that coupling loss vanishes for insulated filaments for all $L$ no matter how large. But in a real conductor, since the filaments are all joined together at the ends of the winding by their low resistance contacts to the current source, at least one twist is needed to zero out the net flux threading the conductor; in practice as many twists as possible should be introduced. This line of reasoning leads us to a twisted group of insulated filaments with no coupling loss. The price for this outcome is an absence of current sharing; the alternative is obvious.

### 4.6.2 The Composite Tape

The composite tape possesses metallic underlayers and overlayers. The underlayer is separated from the YBCO by an insulating buffer layer which can only weakly couple the filaments (see below). On the other hand, the overlayer usually of pure Ag may be in direct contact with the YBCO filaments, and hence will couple them very strongly unless they are striped along with the YBCO. Thus the Ag coating must be striped along with the filaments themselves.

### *4.7 Reduction of Coupling Loss – Further Practical Details*

### *4.7.1 Field Oscillations*

It should be kept in mind that it is possible to reduce eddy current coupling significantly by reversing the field direction within a given length of conductor. In any case, such oscillations must be taken into account when calculating the actual AC loss of a given configuration. In general, if a segment of multifilamentary conductor is exposed to spatially varying fields, it is the net flux through the segment that will determine the effective active length. While we are unaware of any work specifically related to this in the context of coated conductors, the effect is well known in cables [26-28] and LTSC strands themselves [29].

### *4.7.2 Interaction Effects and Calculation Limitations*

The presence of significant oscillating field complicates the detailed calculation of the losses. However, for a first-order estimate, we can in fact expect the dynamic resistance effect to suppress the transport current loss substantially, and hence we need consider only the external magnetic field losses. Some rough estimates for these effects (based on round wire models in various limits) were given above in Section 4.4.3. However, further refinement of these effects are in order.

### *4.7.3 Effective Interfilamentary Resistivity of the Composite Tape*

We consider a model striped composite coated conductor of width $w$ in which the interfilamentary spacing is taken to be the same as the filament width, $d$. The particular choice was made for our model calculations and for comparisons to test samples; in fact the results will be generally valid, even as the interfilamentary spacings get small. The YBCO is deposited on a buffer layer of thickness $t_{buf}$ and resistivity, $\rho_{buf}$. The substrate is of thickness $t_{sub}$ (later $t_{sub} \cong d$) and resistivity $\rho_{sub}$. We will show that although $\rho_{buf}$ is generally very high its contribution to $\rho_{eff}$ is reduced by a factor $2(t_{buf}/w)$, which in practice could be of order $10^{-4}$. The analysis proceeds as follows:

The resistance (per unit length of conductor) of a current path from one outside filament to the other including two passages through the buffer layer is:

$$R_l = \rho_{sub}\left(\frac{w}{t_{sub}}\right) + 2\rho_{buf}\left(\frac{t_{buf}}{d}\right) \tag{7}$$

The effective resitivity is of course

$$\rho_{eff} = R_l \frac{t_{sub}}{w} \tag{8}$$

which becomes

$$\rho_{eff} = \rho_{sub} + 2\rho_{buff} \frac{t_{buf}}{d} \frac{t_{sub}}{w} \tag{9}$$

and with $t_{sub} \cong d$ becomes

$$\rho_{eff} = \rho_{sub} + 2\rho_{buf} \left( \frac{t_{buf}}{w} \right) \tag{10}$$

The lowered effective resistivity is a natural consequence of the fact that the high buffer layer resistivity is localized and "widely separated" by the width of the strand. The effect of the buffer layer in this case is in some ways analogous to that of the resistive barrier that surrounds the filaments in some LTS and HTS MF conductors

### 4.7.4 Loss in the Normal Metal Components of a Composite Tape Conductor

#### 4.7.4.1 Ferromagnetic Loss

A requirement of the RABiTS process is the need for a textured Ni or Ni-alloy substrate with the accompanying ferromagnetic-hysteresis loss. The loss per unit volume of a ferromagnetic substrate is given by [30]

$$P_{Fe} = 4\mu_0 H_c M_s f \tag{11}$$

where $H_c$ is the coercive field and $M_s$ is the saturation magnetization. For a Ni substrate with $M_s = 485$ kA/m and a published $\mu_0 H_c$ of 7 x $10^{-5}$ T [31], we find $P$ (at $f = 500$ Hz) to be 6.8 x $10^4$ W/m$^3$. This leads to a hysteretic loss in a Ni strip 1 cm wide and 50 µm thick of 34 mW/m. In practice a textured and slightly work-hardened Ni strip has been stated to have a measured $\mu_0 H_c$ of 7 x $10^{-4}$ T [5] resulting in a loss that is ten times greater, about 340 mW/m. The ferromagnetic-hysteresis loss can be reduced by alloying the Ni with W or Cr which reduces both $M_s$ and $H_c$. For example, use of Ni-7 at.%Cr with an $M_s$ of 195 kA/m and a $\mu_0 H_c$ of 4 x $10^{-4}$ T reduces the loss by a factor 4. Under the above conditions such a ribbon would dissipate a ferromagnetic loss of 78 mW/m. A better choice (from the standpoints of both magnetic properties and buffer-layer compatibility [32]) would be a Ni-W alloy, e.g. Ni-5 at.%W with hysteretic loss (when annealed at 1,000°C ) of about 270 J/m$^3$-cycle [32] and hence an alloy-strip loss of 67 mW/m.

How significant is this 67 mW/m compared to the composite's other loss components? It is certainly small compared to the FO loss of a model striped conductor (2.5 W/m for 50 µm wide stripes) but comparable to its pure EO loss (100 mW/m, or 50% of the unstriped value due to the correspondingly lower fraction of superconductor present). Furthermore, as noted above, a significant loss contribution by a nominally EO-oriented tape is expected to be the stray (FO) field component. At an amplitude equal to 2% of the EO $H_m$ (e.g. 2% of 1 T in this case) the stray field would contribute (2/100)x(1/2)x100 mW/m = 50 mW/m of loss (c.f. Section 4.3.2). The total "EO loss" in this case would therefore be 150 mW/m. Thus the ferromagnetic loss in the alloyed Ni, at a little less than ½ of the total EO loss, must be regarded as significant and hence a candidate for elimination.

### 4.7.4.2 Eddy Current Loss in Normal Metal Underlayers and Overlayers

Eddy current loss can be expected from the two normal-metal components of the composite ribbon -- the underlayer (substrate) and the overlayer(s). This loss, on a volumetric basis, is given by [30]:

$$P_e = \frac{\pi^2}{6\rho_n}[(\mu_0 H_m)wf]^2 \tag{12}$$

It is useful to compare this with the coupling loss as in

$$\frac{P_e}{P_{coup}} = \left(\frac{2}{3}\right)\left(\frac{\pi}{L}\right)^2\left(\frac{\rho_{eff}}{\rho_n}\right)w^2 \tag{13}$$

Since, as already indicated, $\rho_{eff}$ can be very much larger than any normal-metal resistivity, $\rho_n$, the eddy current loss $P_e$ can be greater than $P_{coup}$ unless the underlayer and overlayer contributions can be reduced by each of two approaches. (1) The $\rho_n$ of the underlayer should be increased. (2) Since the $\rho_n$ of the overlayer(s) will generally be low, their eddy current loss contribution (again, per unit volume) can be reduced by striping them along with the superconductor; in other words by replacing $w^2$ by $d^2$. Thus subdivision into 100 50 µm-wide stripes would reduce the unit-volume overlayer loss by a factor 4 x $10^4$. This reduces the pure-Ag full width (1 cm) FO eddy-current loss of 425 W/m down to 5.3 mW/m after allowing for the gaps between the stripes.

*The Underlayer:* Underlayers for the RABiTS process are Ni alloys. Likewise, that for the IBAD process is usually a Ni alloy (Hastelloy type). At a resistivity 0.6 µΩcm at 77 K a pure Ni substrate of standard dimensions (1 cm x 50 µm) would contribute an eddy current loss at $f$ = 500 Hz, $H_m$ = 1 T of 3.4 kW/m. This unacceptably large loss can be significantly reduced by alloying with Cr or preferably W (RABiTS) or about 30 wt.% Mo or Cr+Mo (Hastelloy, IBAD). The resistivities of these Ni alloys are an almost temperature independent 50-125 µΩcm [33][34]. Even so such alloys at a nominal resistivity of 100 µΩcm would provide eddy current losses of about 20 W/m. But this is still an order of magnitude greater than the FO hysteretic loss of the striped conductor (2.5 W/m). Taken together, these results again emphasize the importance of arranging for an EO orientation of the applied field.

*The Overlayer:* The primary overlayer is a protective film of Ag (77 K resistivity 2.9 x $10^{-7}$ Ωcm [35]). At a thickness of 3 µm, and if insulated from the YBCO, this would contribute 425 W/m to the total loss, 20 times greater than that of the Ni-alloy underlayer. Apart from the "EO solution" there are two ways around this problem: (1) Alloy the Ag with Au to raise the 77 K resistivity into the range of 3.88 x $10^{-6}$ Ωcm [35,36] and reduce the loss to 32 W/m, comparable to that of the Ni-alloy underlayer. (2) Place the Ag film in intimate electrical contact with the YBCO layer to completely eliminate the eddy current. By the same token a superposed pure Cu "stabilization" layer in electrical contact with the Ag and hence the YBCO will not contribute to the eddy current loss. Assuming that we have already agreed upon the necessity of striping the YBCO coating, the overlayer must not be permitted to connect the filaments. The result of so doing would be very large coupling loss. To prevent this, the overlayer should be striped along with the YBCO coating.

### 4.8 Implications for the Conductor and Loss Mitigation

### 4.8.1 Eddy Current, Coupling, and Hysteretic Losses in the Composite Tape

We have shown the need for striping the YBCO layer in order to control the FO hysteretic loss. Overlayers applied *after* the striping fill the interfilamentary spaces with high conductivity metal and re-introduce large coupling loss. But striping performed after a well-bonded coating has been applied maintains the benefit of zero eddy current loss in the overlayer and reduced FO hysteretic loss in the superconductor. An underlayer alloy having been chosen, apart from thickness reduction, nothing can be done to reduce its eddy current loss per meter; it is insulated from the YBCO and is unable to be striped.

### 4.8.2 Ferromagnetic Shielding

As pointed out in connection with research on Fe-clad $MgB_2$ strands some interesting loss properties accompany the shielding of a superconductor by a thin layer of Fe [37-39]. The hysteretic loss is in some cases reduced, and the field lines through the superconductor may be modified by the presence of the Fe. But a *fully shielded* superconductor will not be able to perform its function in an electromagnetic device. In the interest of loss minimization, EO operation is desirable but not necessarily fully attainable in practice, as explained in Section 4.3.2. Even a one degree misorientation of the EO field would activate some FO loss to the extent of some 100 times the pure EO value. This could to some extent be alleviated by suitably engineering an FE layer to redirect the stray field closer to the EO direction.

### 4.8.3 Coated Composite Tapes for AC Applications – Design Summary

First we need to choose a non-magnetic substrate with sufficiently high resistance. Beyond this, we must be careful with our choice of buffer layers, such that the interfilamentary resistance is sufficiently high. Then we must sparingly apply the conductive overlayer, using an alloy such as Ag-Mg. This overlayer should also be striped, along with the YBCO layer. There are several potentially important components to the loss: hysteretic loss of the YBCO coating, coupling loss of the striped composite, ferromagnetic loss in a Ni-alloy underlayer if present, and normal metal eddy current losses in the underlayer and overlayer. For large-$H_m$ applications it will be necessary to introduce filamentary subdivision of the YBCO, to reduce the hysteretic loss. It will then be desirable to use very resistive underlayers and overlayers (excepting in the latter case when a stabilization function is required). It is important to note that the eddy current loss from an unstriped normal metal layer in intimate contact with an unstriped YBCO coating would be low even in FO fields. The same would also be true for a comparable striped conductor provided that the striping were performed after the overlayer had been applied. The converse would lead to large coupling loss if the overlayer were in good electrical contact with the superconductor, or large eddy current loss if it were insulated from it, e.g., by a buffer layer.

## 5. DISCUSSION

Table 4 lists various loss contributions for a number of coated conductor designs and orientations, all with striped filaments and overlayers. In all cases $I_c = 200$ A before striping, and 100 A after striping. The underlayer is 50 μm thick Hastelloy-type (100 μΩcm). The overlayer is 3 μm thick Ag, and assumed to be striped along with the filaments. The striping is assumed to remove 50% of the filamentary region. This is too pessimistic for 50 μm filaments -- even now better superconducting fractions can be achieved and this will continue to improve. However, a 50% superconducting fraction it is probably realistic (for now) for 5 μm filaments and so for consistency of comparison we choose this value. All values are calculated for 77 K operation. In generating the data for Table 4 we have used the expressions from Section 4, above. Thus, Table 4 gives a first order approximation (estimate) of the values for each of these loss components. Hysteretic loss is calculated using the field independent $J_c$. Normal metal eddy currents are calculated using Eq. 12 directly, thus any suppression of this loss in the overlayer due to its close connection to the YBCO ( caused by electric field suppression due to physical proximity and high electrical connectivity to the overlayer) is ignored. Based on our discussions above regarding the futility of normal-metal-connection-enabled current sharing at 500 Hz, the filaments are assumed to be connected to one another only via the buffer layers to the substrate. In addition, the influence of end-connections on $P_c$ is ignored (a point of view valid for a long twisted sample); taken together with the highly resistive buffer layer connections this makes the coupling term negligible. Ferromagnetic loss, $P_{Fe}$ is calculated for a Ni-5%W underlayer, and can be deleted for Hastelloy or other non-magnetic underlayers (e.g., higher Cr content Ni-Cr alloys). Transport current losses are also shown in Table 4. Here row (**3a**) assumes no applied field. The real contribution of this term will be less than that shown as the fields become large enough to decouple the filaments with respect to their self field. We have estimated those results for the regions of the conductor experiencing large fields (from our discussions above) and they are listed in row (**3b**). Below, the data of Table 4 are discussed with reference to the stator and rotor requirements of an all-superconducting generator operating at 500 Hz..

*The Stator:* Cryogenic Cu machines operating at liquid hydrogen temperatures would have conductors which, once scaled to a 200 A level, would generate roughly 1.5-3 W/m. While this would be acceptable for an open cycle machine, a closed cycle machine would demand less, certainly below 500 mW/m. Our reference design (one simplified, but basic potential design [22]) for the superconducting stator calls for an AC field amplitude of 1 T. Some designs for *each phase* of a stator call for 150 m of 1000 A conductor, and a loss budget of 100 W. This means 750 m of the "standard" 200 A conductor, with **130 mW/m as its loss target**

*The Rotor:* For a rotor, the situation is different.. Fortunately, the AC fields to which it is exposed are relatively small. The rotor is a DC magnet which experiences just a ripple field by inductive coupling to the stator – e.g. with an amplitude of only 2% of the stator field amplitude (assumed to be 1 T). Of this only about 10% would be FO oriented. Thus on this basis the FO field experienced by the rotor has an $H_m$ of about 2 mT. It is estimated that for the rotor some 4000 m of 150 A wire would be needed, with a loss budget of **0.5 mW/m as the target.**

With these targets in mind we are now in a position to examine Table 4 in some detail. Consider first the stator requirements. Several possible conductor geometries are listed, denoted (**1a**), (**1b**), and (**1c**). Geometry (**1a**) is a 1 cm wide tape with 50 μm filaments. Here the losses for FO oriented fields (as might be expected for stator coils using a Gramme ring geometry) are far too high. In going to (**1b**) we reduce the filament width by an order of magnitude. This reduces the hysteretic component, but the substrate losses are still much too high. Only by going to (**1c**) can we reduce the loss to 450 mW/m which is getting close to the goal. But we have not yet taken into account the duty cycle of the winding. We have (as a first order approximation assumed that all sections of the Gramme ring conductor would be experiencing the 1 T AC field at all times. The loss requirement needs to be down-rated because: (1) only the inner portions of the winding (those nearest the rotor) experience the full field, and (2) only some fraction of the Gramme ring winding experience the full field at any given time in the cycle. The reduction factor associated with these effects (unpublished to our knowledge) might result in a factor of two for each effect separately. The final loss estimate of 112 mW/m puts the Gramme ring stator, wound with Option-(**1c**) conductor, within range of the target. The Gramme-ring might be regarded as a "worst-case scenario". That winding geometry can be used as a means to keep the bending radius of the conductor as large as possible– but it can be argued that bending the conductor is less damaging than twisting it. Indeed, YBCO coated conductor has been demonstrated to have almost double the strain tolerance of BSCCO HTS conductor and can tolerate a smaller bend radius [40]. A simple conception of twisting, on the other hand, would tend to reduce $J_e$ considerably, and might be difficult to do in practice for  any device winding. If, alternatively, we go to a more traditional diamond-geometry stator it should be possible to avoid twisting and to additionally align the conductors mostly EO. At 2% stray field they would be exposed to an equivalent FO field amplitude of  0.02 T; allowing for somewhat greater misorientation we might consider a larger $\mu_0 H_m$, somewhere between 0.02 and 0.2 T and hence a specific loss close to the 130 mW/m target and with a more easily fabricable conductor geometry – Option (**1b**).

As for the rotor, which would be exposed FO to an AC field component of only 2 mT,  conductor (**1c**) will satisfy the 0.5 mW/m loss requirement. Note that we have assumed no shielding for the rotor. A shield can be included, in which case the loss is transferred to it in the form of eddy currents.

# 6. CONCLUDING SUMMARY

AC losses for coated conductors have been investigated by calculation and experiment. Based on previous studies the utility of striping as a way of reducing hysteretic loss was verified. A suggested YBCO striping geometry which includes striping the stabilizer was described. Such striping was then performed on segments of coated conductor using laser micromachining. Various machining parameters were investigated, and the stripe and slicing characteristics were presented. Typical stripe and slice widths were 25 μm at the level of the YBCO layer, and the depth of the slice could be controlled, sometimes through the use of multiple passes. Some simple striping patterns were selected and packs of coated conductor (some including strips of Fe for partial shielding) were prepared for AC loss measurement. For the 10-filament samples measured the loss reduction was consistent with the factor of 10 expected. Fe shielding lowered the loss for unstriped samples, but increased the loss of striped samples (or rather, striped YBCO/Fe sample packs had greater loss than those without Fe). An eddy current component was seen only for the samples with Fe; this contribution was the normal metal eddy current within the Fe itself. In general, striping was shown to be very effective for loss reduction, while ferromagnetic shielding was only partially effective. Based on the outcome of what might be regarded as preliminary experiments at AC field frequencies of 50 to 200 Hz and amplitudes of up to 150 mT detailed calculations were made of losses at 500 Hz and 1 T. Then by way of a Conductor Design exercise, values were estimated for various components of the total loss, including hysteretic, normal metal eddy current, coupling eddy current, and transport loss. In the high frequency/moderate field regime the dominant contributions are shown to be hysteretic losses and eddy currents in FO applied fields. The conductor design study concluded with a tabulation of loss contributions from all anticipated sources for two frequencies (200 and 500 Hz) and several AC field amplitudes (1, 0.2, 0.02, and 0.002 T) and an indication as to how the various cells of the table describe conductors for use in particular large-scale applications.

## Acknowledgements

We thank the supplier for the YBCO coated conductor and Ken Hix from Mound Laser Photonics for laser striping these samples. We also appreciate helpful discussions with Larry Long of LEI. This research was funded by an AFOSR STTR (via Hyper Tech Research) and an AFSOR summer faculty program.

List of Tables

Table 1. Striping Parameters.

| Stripe Name | Speed, mm/min | Width at Ag, μm | Width at YBCO, μm | Depth, μm | Beam Power at tape, W | Atm | No. pass |
|---|---|---|---|---|---|---|---|
| G | 250 | 36 | 28.70 | 37.60 | 0.75 | Air | 1 |
| F | 300 | 24 | 23.65 | 33.00 | 0.75 | Air | 1 |
| E | 350 | 34 | 27.69 | 28.10 | 0.75 | Air | 1 |
| D | 400 | 39 | 30.95 | 31.59 | 0.75 | Air | 1 |
| C | 450 | 30 | 30.16 | 27.93 | 0.75 | Air | 1 |
| B | 500 | 36 | 29.04 | 31.60 | 0.75 | Air | 1 |
| M | 250 | 29 | 26.82 | 36.82 | 0.50 | Ar | 1 |
| L | 300 | 32 | 26.65 | 36.02 | 0.50 | Ar | 1 |
| K | 350 | 35 | 26.99 | 30.63 | 0.50 | Ar | 1 |
| J | 400 | 34 | 23.48 | 28.87 | 0.50 | Ar | 1 |
| I | 450 | 30 | 26.67 | 29.90 | 0.50 | Ar | 1 |
| H | 500 | 35 | 24.28 | 31.90 | 0.50 | Ar | 1 |
| S | 250 | 31 | 25.08 | 49.03 | 0.50 | Air | 1 |
| R | 300 | 39 | 25.08 | 42.23 | 0.50 | Air | 1 |
| Q | 350 | 32 | 21.97 | 38.57 | 0.50 | Air | 1 |
| P | 400 | 36 | 27.14 | 36.83 | 0.50 | Air | 1 |
| O | 450 | 34 | 23.95 | 31.59 | 0.50 | Air | 1 |
| N | 500 | 34 | 21.27 | 26.97 | 0.50 | Air | 1 |
| T | 550 | 33 | 23.65 | 31.27 | 0.50 | Air | 1 |
| U | 600 | 32 | 26.18 | 29.04 | 0.50 | Air | 1 |
| V | 650 | 39 | 30.46 | 30.78 | 0.50 | Air | 1 |

Table 2.  Slicing Parameters.

| Slice Name | Speed, mm/min | Width at Ag, μm | Width at YBCO, μm | Depth, μm | Beam power at tape, W | Atm | No. pass |
|---|---|---|---|---|---|---|---|
| W | 700 | 27.46 | 23.65 | 31.89 | 0.50 | Air | 1 |
| X | 700 | 26.92 | 22.80 | 30.16 | 0.50 | Air | 2 |
| Y | 700 | 36.83 | 28.73 | 50.15 | 0.50 | Air | 5 |
| Z | 700 | 33.99 | 27.61 | 88.87 | 0.50 | Air | 10 |
| ZA | 700 | 37.45 | 29.83 | 118.0 | 0.50 | Air | 15 |
| ZB | 700 | 58.09 | 51.75 | 117.8 | 0.50 | Air | 20 |
| ZC | 700 | 53.64 | 47.30 | 132.1 | 0.50 | Air | 30 |
| ZD | 700 | 50.17 | 46.66 | 132.0 | 0.50 | Air | 40 |

Table 3. Sample Pack Specifications.

| Sample Pack | Striped YBCO | Unstriped YBCO | Fe strips | Total length YBCO, m | Fe/Tape Vol ratio |
|---|---|---|---|---|---|
| A | -- | 6 | 0 | 0.3 | 0.0 |
| B | -- | 6 | 4 | 0.3 | 0.063 |
| C | 2 | -- | 0 | 0.1 | 0.0 |
| D | 2 | -- | 2 | 0.1 | 0.167 |

Table 4. AC Loss Component Estimates Under Various Conditions.

| Condition/ Loss in mW/m | | 500 Hz, 1 T | 500 Hz, 0.2 T | 500 Hz, 0.02 T | 500 Hz, 0.002 T | 200 Hz, 1 T | 200 Hz, 0.2 T |
|---|---|---|---|---|---|---|---|
| **(1a)** | $P_h{}^a$ | 2500 | 500 | 50 | 5 | 1000 | 200 |
| Face-on | $P_{e,sub}{}^b$ | 20,000 | 800 | 8 | 0.08 | 3,200 | 128 |
| $w$ = 1 cm | $P_{e,over}{}^c$ | 5.3 | 0.212 | .002 | -- | 0.85 | 0.034 |
| $d_f$ = 50 μm | $P_c{}^d$ | -- | -- | -- | -- | -- | -- |
| | $P_{Fe}{}^e$ | 68 | 68 | ?? | ? | 27 | 27 |
| **(1b)** | $P_h$ | 250 | 50 | 5 | 0.5 | 100 | 20 |
| Face-on, | $P_{e,sub}$ | 20,000 | 800 | 8 | 0.08 | 3,200 | 128 |
| $w$ = 1 cm | $P_{e,over}$ | 0.053 | 0.002 | -- | -- | 0.009 | -- |
| $d_f$ = 5 μm | $P_c$ | -- | -- | -- | -- | -- | -- |
| | $P_{Fe}$ | 68 | 68 | ?? | ? | 27 | 27 |
| **(1c)** | $P_h$ | 250 | 50 | 5 | 0.5 | 100 | 20 |
| Face-on | $P_{e,sub}$ | 200 | 8 | 0.08 | 0.001 | 32 | 1.28 |
| 10 strips | $P_{e,over}$ | 0.053 | 0.002 | -- | -- | 0.009 | -- |
| $w$ = 1 mm | $P_c$ | -- | -- | -- | -- | -- | -- |
| $d_f$ = 5 μm | $P_{Fe}$ | 68 | 68 | ?? | ? | 27 | 27 |
| **(2)** | $P_h$ | 100 | 20 | 2 | 0.02 | 40 | 8 |
| | $P_{e,sub}$ | 0.5 | 0.02 | -- | -- | 0.08 | 0.003 |
| Edge-on | $P_{e,over}$ | 0.019 | 0.001 | -- | -- | 0.003 | -- |
| | $P_c$ | -- | -- | -- | -- | -- | -- |
| | $P_{Fe}$ | 68 | 68 | ?? | ? | 27 | 27 |
| **(3a)** | $P_{sf}(1a,b)$ | 88 | 88 | 88 | 88 | 35 | 35 |
| AC current/ self field $i$=0.5 | $P_{sf}(1c)$ | 8.8 | 8.8 | 8.8 | 8.8 | 3.5 | 3.5 |
| **(3a)** | $P_i(1a)$ | 208 | 41.6 | 4.16 | 0.416 | 83 | 16.6 |
| AC current/ | $P_i(1b)$ | 20.8 | 4.16 | 0.416 | 0.042 | 8.3 | 1.66 |
| AC field FO $i$=0.5 | $P_i(1c)$ | 20.8 | 4.16 | 4.16 | 0.416 | 8.3 | 1.66 |

[a] $P_h$ = Superconductor hysteretic loss (YBCO layer, 2 μm thick). No $N$-factor applied here.

[b] $P_{e,sub}$ = Normal metal eddy current loss for substrate (Hastelloy-type alloy, 50 μm thick, 1 x $10^{-6}$ Ωm).

[c] $P_{e,over}$ = Normal metal eddy current loss for overlayer (Pure Ag, 3 μm thick, 2.9 x $10^{-9}$ Ωm. The overlayer is assumed to be striped. Values do not include the influence of the SC on the electric field).

[d] $P_c$ = Coupling eddy current loss.

[e] $P_{Fe}$ = Ferromagnetic substrate hysteretic loss (Ni-5%W here, but could be zero if nonmagnetic substrate was used).

$P_{sf}$ = Transport current loss, self field.

$P_i$ = Transport current loss, in presence of AC field.

List of Figures

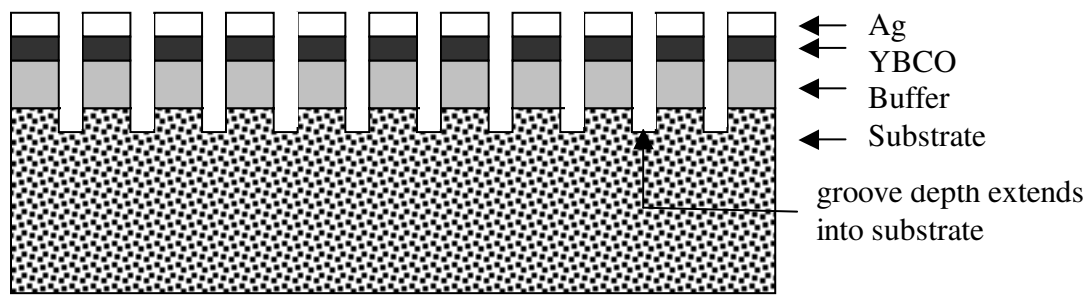

Ag
YBCO
Buffer
Substrate

groove depth extends
into substrate

Figure 1

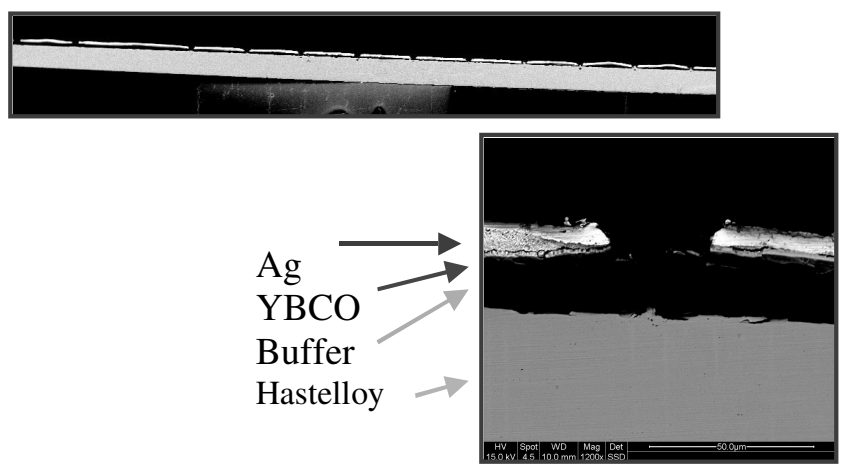

Figure 2.

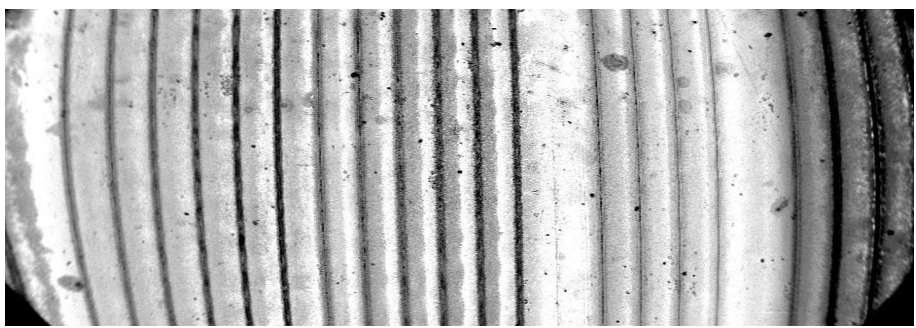

Figure 3.

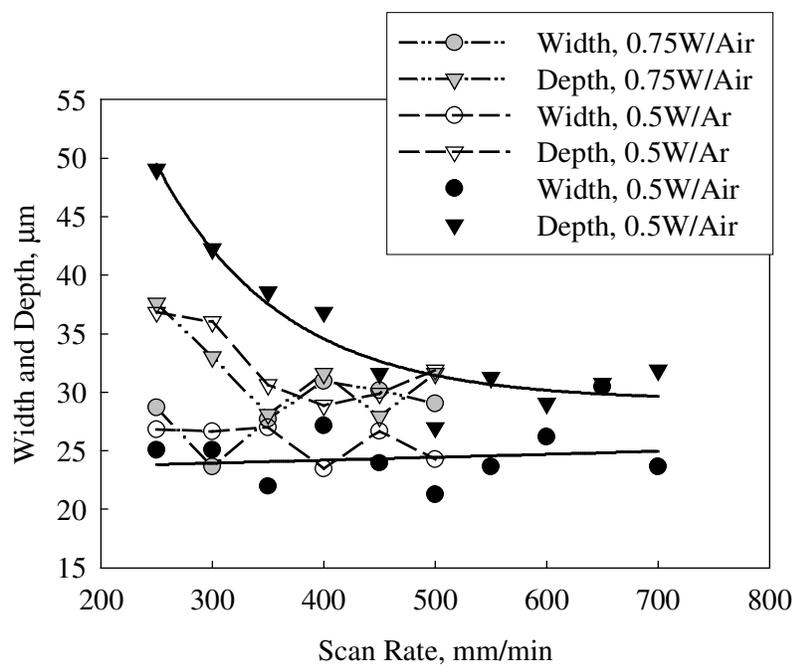

Figure 4.

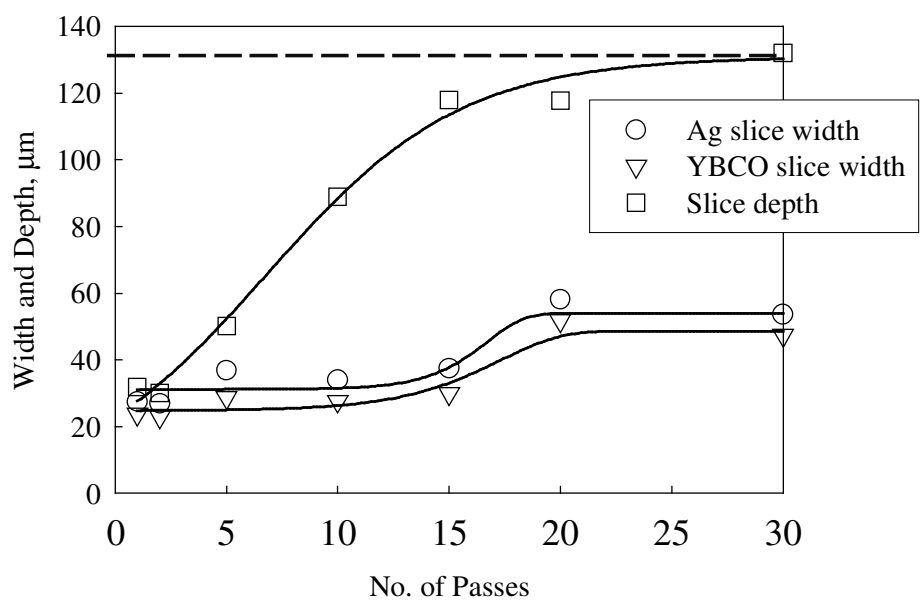

Figure 5.

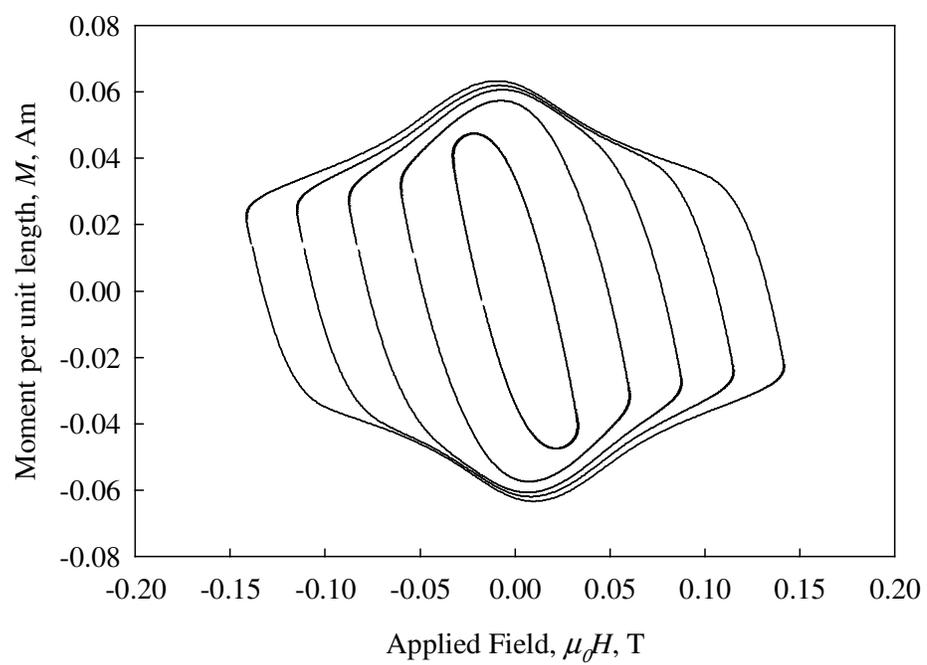

Figure 6.

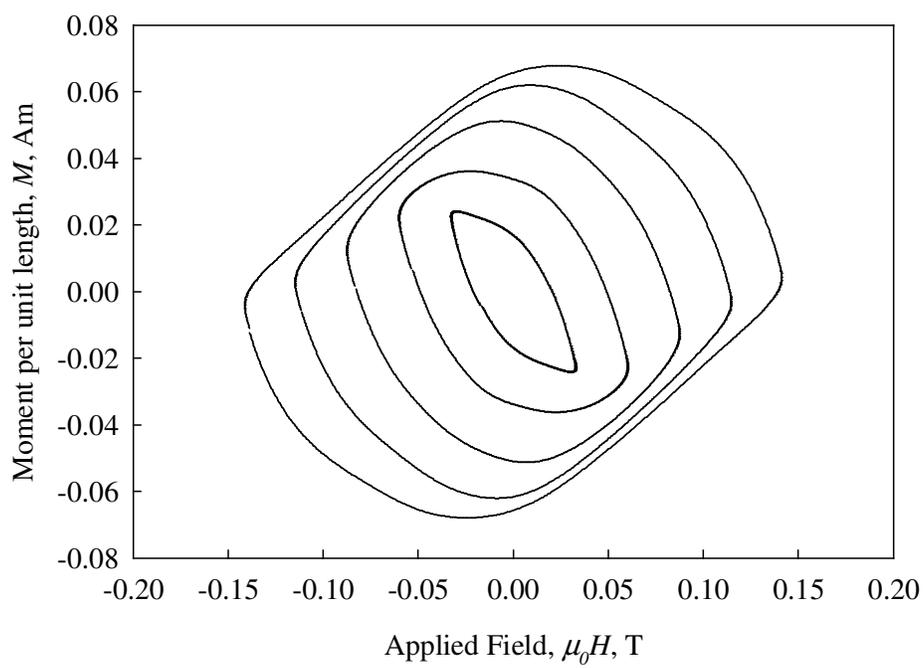

Figure 7.

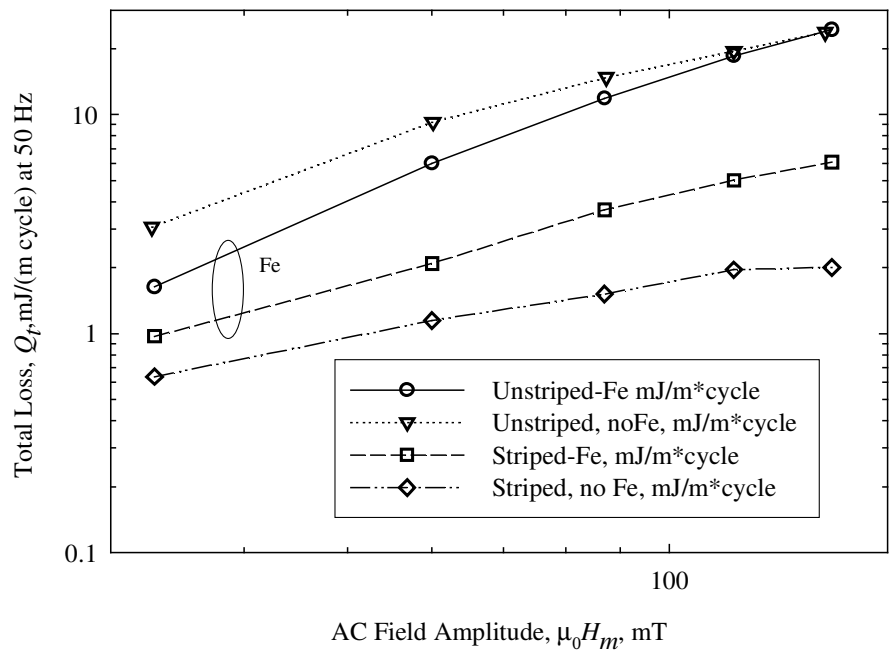

Figure 8.

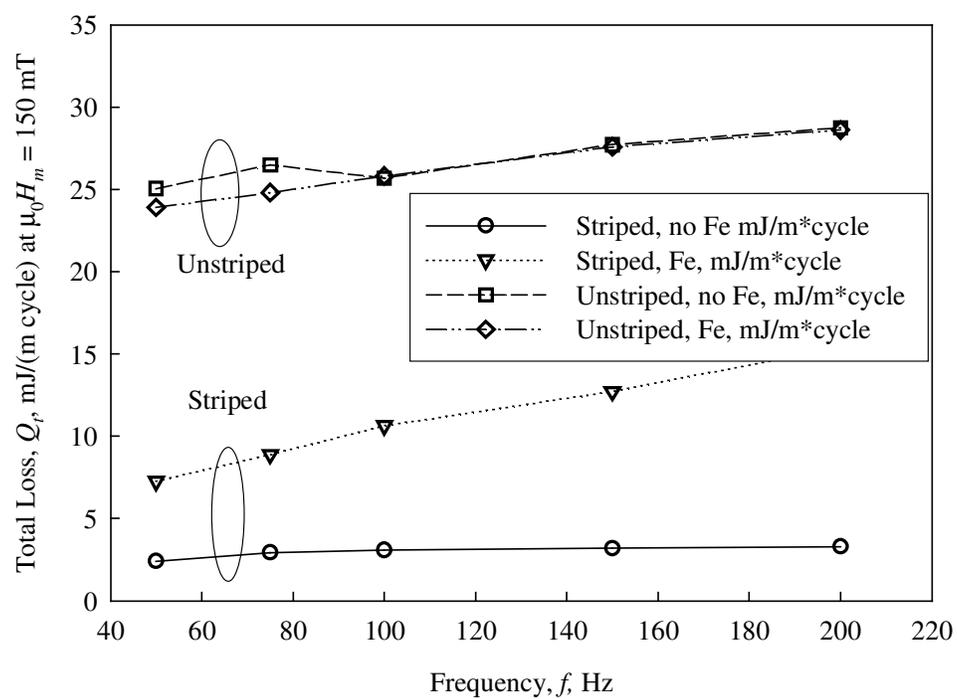